# A FRAMEWORK STUDIO FOR COMPONENT REUSABILITY

N Md Jubair Basha[1] and Salman Abdul Moiz[2]

[1]Information Technology, Muffakham Jah College of Engineering & Technology, Hyderabad, INDIA
`jubairbasha@mjcollege.ac.in`
[2]Professor CSE/IT, MVSR Engineering College, Hyderabad, INDIA
`Salman.abdul.moiz@ieee.org`

*ABSTRACT*

*The deployment of a software product requires considerable amount of time and effort. In order to increase the productivity of the software products, reusability strategies were proposed in the literature. However effective reuse is still a challenging issue. This paper presents a framework studio for effective components reusability which provides the selection of components from framework studio and generation of source code based on stakeholders needs. The framework studio is implemented using swings which are integrated onto the Net Beans IDE which help in faster generation of the source code.*

*KEYWORDS*

*Component, Reusability, Framework, Product*

## 1. INTRODUCTION

Effective software reuse helps in development of quality product within time and budget. This also helps in reducing the high effort needed for testing and maintenance of the software products.

Several tools are proposed in literature which provides only subset of operation requirements of effective software reuse. Though some of the tools generate source code as an artifact, but they may not work for enterprise software. In this paper a framework for effective software reuse is presented which is used to identify the components of a given enterprise requirements. The identified components are integrated onto the Net Beans IDE to generate a deployable application.

The remaining part of this paper is organized as follows: section-2 presents the advantages of reusing software systems, section-3 describes approaches and tools of domain engineering, section-4 describes the proposed component framework along with the process followed for effective code generation, section -5 shows the simulation results and section -6 concludes the paper.

## 2. SOFTWARE REUSE

Software Reuse is the use of available software or to build new software from software knowledge. Reusable assets can be either reusable software or software knowledge. Reusability is a property of a software asset that indicates it's probability of reuse [1]. Software Reuse means the process that use "designed software for reuse" again and again [2]. By software reusing, we can manage complexity of software development, increase product quality and makes faster production in the organization.

Recently, design reuse has become popular with (object-oriented) class libraries, application frameworks, design patterns and along with the source code [3]. Jianli et al. proposed two complementary methods for reusing existing components. Among them one allows component evolution itself, which is achieved with binary class level inheritance across component modules. The other is by defined semantic entity so that they can be assembled at compile time or bind at runtime. Although component containment still is the main reuse model that leads to contribute the software product lines development [4]. Regarding the components much information has to be collected, maintained and processed for the retrieval of the components. Maurizio has proposed a methodology to automatically build a software catalogue the tools for archiving and retrieval of information are presented [5]. Software Reuse can be broadly divided into two categories viz. product reuse and process reuse. The product reuse includes the reuse of a software component and by producing a new component as a result of module integration and construction. The process reuse represents the reuse of legacy component from repository. These components may be either directly reused or may need a minor modification. The modified software component can be archived by versioning these components. The components may be classified and selected depending on the required domain. [6].

## 3. DOMAIN ENGINEERING

Software Reuse can be improved by identifying objects and operations for a class of similar systems, i.e. for a particular domain. In the context of software engineering domains are application areas [7].

There are various definitions of what a *domain* is. Czarnecki's defines [8]:" an area of knowledge scoped to maximize the satisfaction of the requirements of stakeholders, which includes concepts and terminology understood by practitioners in the area and the knowledge of how to build (part of) systems in the area".
Domain Engineering is a process in which the reusable component is developed and organized and in which the architecture meeting requirements of the domain is designed [9].

Domain Engineering can be defined by identifying the candidate domains and performing domain analysis and domain implementation which includes both application engineering and component engineering. Domain Analysis is a continuing process of creating and maintaining the reuse infrastructure in a certain domain. The main objective of domain analysis is to make the whole information readily available. The relevant components (if available) has to be extracted from the repository rather than building the new components from the scratch for a particular domain.

Domain Analysis mainly focuses on reusability of analysis and design, but not code.This can be achieved by building common architectures, generic models or specialized languages that additionally improve the software development process in the specific problem area of the domain. A vertical domain is a specific class of systems. A horizontal domain contains general software parts being used across multiple vertical domains. Mathematical functions libraries container classes and UNIX tools are the examples of horizontal reuse. The purpose of domain engineering is to identify objects and operations of a class in a particular problem domain [7].

In the process of domain analysis, each component identified can be categorized as follows.

- General-purpose components: These components can be used in various applications of different domains (horizontal reuse).
- Domain-specific components: They are more specific and can be used in various applications of one domain (vertical reuse).
- Product-specific components: They are very specific and custom-built for a certain application, they are not reusable or only useful to a small extent.

## 3.1. APPROACHES FOR DOMAIN ENGINEERING

Several tools for domain engineering are proposed in literature. Each tool specifies subset of operational requirements needed for the realization of domain engineering.

• Domain Analysis and Reuse Environment (DARE-1988) is a tool that provides a mechanism to capture the information needed from source code, documentation and from domain experts. The extracted domain information is stored in a repository that typically contains a generic architecture for the domain and domain-specific reusable components. For the purpose of reusing the available domain information from repository, DARE provides a library search facility to retrieve the stored domain information [10].

• Family-Oriented Abstraction, Specification and Translation (FAST) provides API's based on an application modeling language (AML). It helps developers to create the tools that are needed for generating software product line by using domain engineering & application engineering phases.

• Feature Oriented Reuse Method (FORM) is a systematic method for capturing and analyzing the similarities and differences of features pertaining to specific domain. It is an extension to the Feature Oriented Domain Analysis (FODA).  The identified domain architectures and components helps to discover and understand the similarities and differences of a product line [12].

• Kobra (KomponentenbasierteAnwendungsentwicklung) is used for component-based development [1]. This method consists of product line development, component based software development and frameworks to provide systematic approach to developing high quality component based application frameworks [13]. As it can be used with three major component implementation technologies viz., Java Beans, CORBA and COM, it is assumed to be "technology independent".

• Product Line UML-Based Software Engineering (PLUS) is a model-driven evolutionary development approach for software product lines. It is used for analyzing and modeling a single system. Additionally, it provides a set of concepts and techniques to explicitly model the similarities and differences in a software product line. This approach enables object oriented requirements, analysis and design models of software product lines to be developed using UML 2.0[14].

• AndroMDA is an extensible generator framework that adheres to the Model Driven Architecture (MDA) paradigm [15]. AndroMDA presents a systematic development methodology to generate deployable components for several platforms and technologies. This helps in achieving the goal of "write less code". The architecture generated by AndroMDA is relatively standard to modern enterprise applications. It consists of four stacked layers, each one containing several components providing some specific functionality. Each component, represented by an AndroMDA, can communicate with other components of the same layer or directly with layers below it.

• With Poseidon [16] for domain specific languages can quickly create fully graphical editor for any domain, customize the editor according to the user needs and generate output from the models.

• QT is a cross platform for C++. Additionally QT has IDE like Eclipse for C++ on mobile . It allows to write advanced applications and UIs once & deploy them across desktop and embedded operating systems without rewriting the source code. This helps in saving the development and testing effort. Whether it is C++ or JavaScript, QT provides the building blocks — a broad set of customizable widgets, graphics canvas, style engine and modern user interfaces. It also incorporates 3D graphics, multimedia audio or video, isual effects, and animations to deploy the applications. QT provides cross-platform IDE. QT's integration with the WebKit web rendering engine helps in quickly incorporating content and services from the Web into the native application. It can use the web environment to deliver the required services and functionality for the satisfaction of end users [17].

## 4. FRAMEWORK STUDIO

There is a need for tools to support the creation of domain specific collections of reusable components, also known as Framework, which is tuned specifically for a particular application domain, i.e. an interface building framework. A framework consists of set of components that express a design for a family of related applications. Therefore, a framework will not be as generally useful outside the application domain because it contains domain-dependent components. However, it is sometimes beneficial to adopt the developing software so that it fits to a proposed framework, resulting in a tremendous gain in productivity. The graphical features of a CASE environment are proposed and reusing an interactive framework by following a proposed process is presented.

Studio means eclipse and Net beans IDE provides the flexibility to the user to work in the workspace for the development of applications. In J2EE, only pure business logic should be written and it provides security, persistence, internationalization and localization. The proposed Framework Studio is a tool used to create specific application screens for particular domains. This will be used by the developers. Framework code generation component is responsible for code generation using file the produced by the framework studio. This is integrated into the workspace of the Net beans IDE to develop the applications. Framework studio and Framework Code Generation component itself coded to be highly flexible & scalable.

### 4.1 PROS OF FRAMEWORK STUDIO

This Framework Studio enhances the code generation feature. Using the requirements of stake holders, this framework studio can automatically generate the source code depending upon the components identified for a particular domain. This Framework Studio provides the facility to drag and drop the components onto a Canvas from the pallet. This increases the reusability of enterprise applications in particular domain.

　　The advantages of a framework studio are as follows:
- Ease of development
- Uniformity in User Interface development

- Reduce Development Efforts
- Reduce Redundant Work
- Reuse
- Manual and automation intervention is balanced optimally

### 4.2 FRAMEWORK STUDIO PROCESS

The proposed framework can be implemented by following a well defined process. The proposed process for the realization of the given framework is depicted in Figure.1.

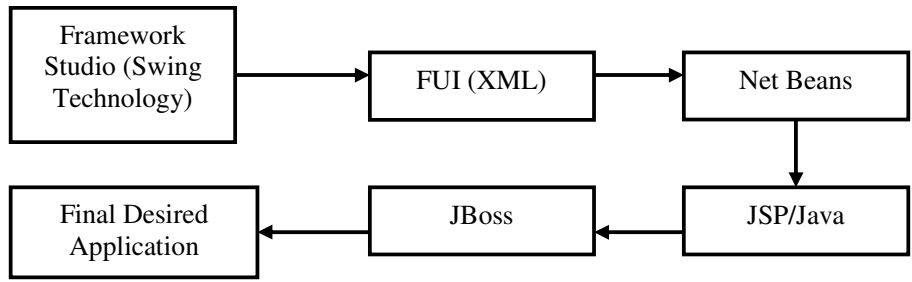

Figure1. Framework Studio Process

The Framework Studio is built using the Swings technology in Java. It represents an integration of many components. These components are bundled on a particular pallet that helps in selection of components of a given domain requested by the stakeholder. In the second phase, the palette is captured in the framework user interface (FUI) which is a XML file. In the third phase, the XML form of the FUI is integrated into the Net Beans IDE to generate JSP/Java files after starting the JBoss server. In the final phase the JSP's are deployed on JBoss server to deploy the desired application which also includes the business logic along with the web application.

## 5. RESULTS

The experimental setup consisted of a machine running Windows-XP Intel Pentium-IV 2.26 GHz and 1.5 GB Ram. The JBoss server running on the Sun Java development kit 1.6. The application was based on the JBoss application server default settings.

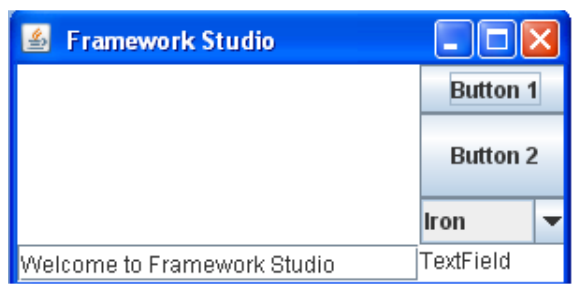

Figure 2. Framework Studio to drag the components present on to the Canvas from Pallet

Framework Studio consists of a Canvas and pallet developed by using the swings technology. The Figure 2, consists of components such as Button 1, Button 2, Combo Box with its list and a Text Field which are available on the Pallet. A Canvas represents a blank rectangular area of the screen onto which the application can draw or from which the application can trap input events from the user. Here, on the Canvas the components which are available on the palette are dragged on to the Canvas. Wherever a selection of component is made, for example, a button or text field etc is selected or dragged on to the Canvas then relevant components size with its name should be added. Likewise, all the components which are needed by the stake holder can be provided with in a mean time without the development of the new components from scratch.

The Net Beans IDE 6.0 is used to run the HR portal application. The HR portal application consists of EAR (Enterprise Archive) file which is a collection of both WAR and JAR files. The WAR (Web Archive) consists of the Servlets, JSP's and HTML web pages. The JAR represents the Java Archive which consists of the business logic classes. The Java EE modules consist of the HR portal-ejb.jar and HR portal-war.war are created.

Under the HR Portal-ejb, the beans available are EmployeeBean, HR Process Bean, InterviewResultsBean classes are created with their methods. The source packages are com.mycompany.hr.dao, com.mycompany.hr.process and com.mycompany.hr.vo were created. In the com.mycompany.hr.dao package, the BaseDAO.java, EmployeeDAO.java, InterviewDAO.java, HRDAO.java and ProcessDAO.java classes are created. The BaseDAO.java is the superclass for the EmployeeDAO.java, InterviewDAO.java, HRDAO.java, ProcessDAO.java subclasses. In the com.mycompany.hr.process package, there are classes such as the EmployeeBean.java, EmployeeBeanRemote.java, HRProcessBean.java, HRProcessBeanRemote.java, InterviewRFesultsBean.java, InterviewRFesultsBeanRemote.java are created.

The business logic methods for the above classes are defined in the developed application. In com.mycompany.hr.vo package there are classes such as CandIntResults.java, CandidateProfile.java, EmpSalary.java, EmployeeCredentials.java, EmployeeProfile.java are created. The other module HR Portal-war.war is created which consists of the Servlets and Jsps. The web pages which were created are Login.jsp, Welcome.jsp, Index.jsp, ViewProfile.jsp, AddCandidate.jsp, InterviewResult.jsp, Registration.jsp. The sourcePackage includes only one package i.e. com.mycompany.hr.servlet package, includes the classes are HRProcessServlet.java, LoginServlet.java, RegistrationServlet.java, InterviewResult.java. This package classes extend the functionality of the HttpServlet superclass.

Thus the two modules i.e. HRPortal-ejb.jar and HRPortal-war.war with their packages, its classes and methods are described.

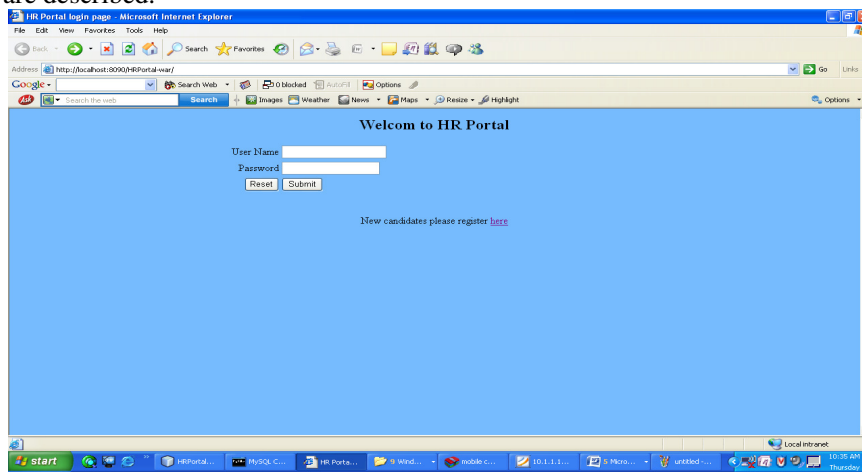

Figure 3. HR Portal Application after integrating into Net Beans IDE

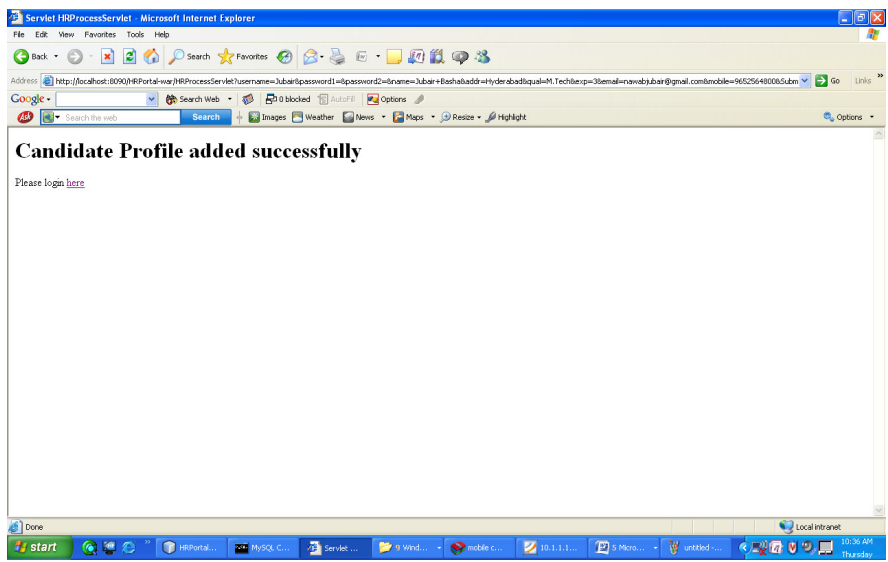

Figure 4. Registration Details Page for HR Portal Application

When the HR Portal application is made to deploy the application by Clean and Build it. At deployment by extracting its JEE Modules i.e. HRPortal-ejb.jar and HRPortal-war.war. The JBoss application server is started. Then the HR Portal application Figure 3 can be seen. If the candidate is already registered to the database then the candidate can directly Sign in the Page which is available in the Figure 3. If the candidate is new candidate, then by clicking on to the here button can be registered. The AddCandidate.jsp page is invoked (see Figure 4.). The AddCandidate.jsp page registers the candidates dynamically. Further, HRProcessServlet.java is invoked. In the invoked servletpage the Login is available. Upon clicking the here button, again the LoginServlet.java is invoked with it's relevant functionality displaying the again the Figure 3. The HRProcessServlet.java is invoked and add the candidates details to the Database via the ProcessDAO.java and BaseDAO.java.Here the getConnection() will invoked to add the candidate details to the database. Further, Figure 5 will be generated.

Figure 5. Servlet Invocation Web Page

Whenever the existing user login form i.e. login.jsp webpage, then the getConnection() will gets invoked. Again Figure 3 allows the existing user to do further more operations.

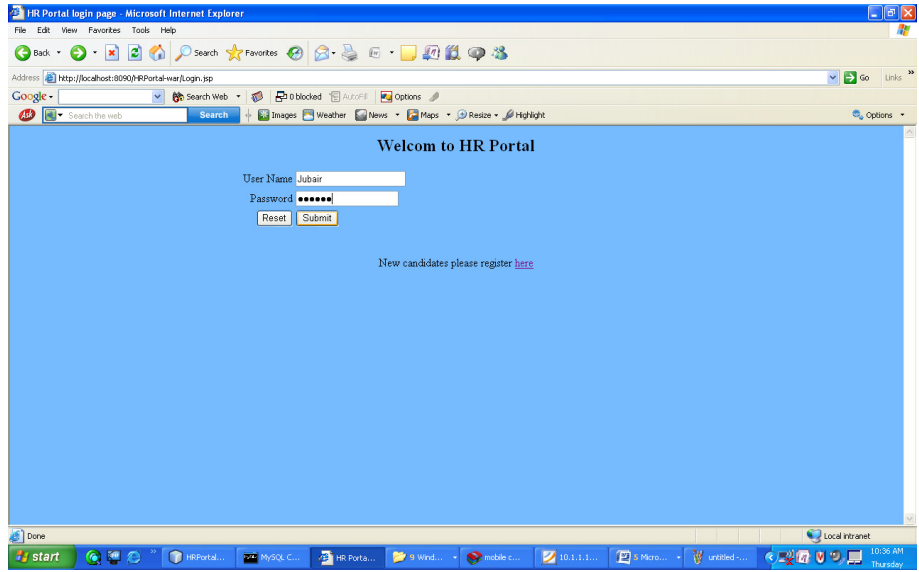

Figure 6. Login Page of the HR Portal Application

Whenever the existing user enters the Username and password (see Figure 6), the processDAO class executes by verifying the authenticate() method so that the relevant username and password which were entered in the login page is there in the database or not. If it has been found, then the LoginServlet.java in conjunct with the HRProcessSevlet.java gets invoked by displaying the webpage Figure 5.

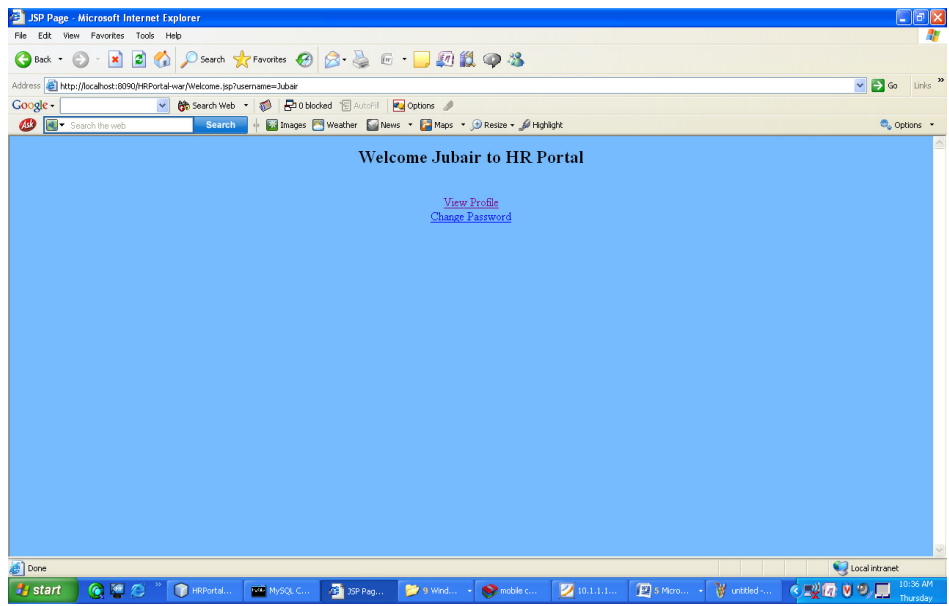

Figure 7. Welcome.jsp page for HR Application

In Figure 7, will invoked when ever the already existing candidate made the login. Again it contains the ViewProfile and ChangePassword. Upon Clicking on to the ViewProfile.jsp will gets displayed.

The database used for this application is My SQL. The tables used in My SQL database which are used in the HR Portal application are Emp_Profile, Emp_Credentials, Emp_Salary, Candidate_Profile, Cand_Int_Results. The Emp_Profile table consists of attributes are emp_id, name, address, dob, experience, doj, email, mobile. The Emp_Credentilals table consists of attributes are emp_id, password. The Emp_Salary table consists of attributes are emp_id, designation, basic, da, hra, cca, pf. The Emp_Profile table consists of attributes are Regn_id, name, address, qual, email, mobile, experience.

In order to test the developed Framework Studio, HR Portal Application is developed without developing of the enterprise application i.e. its business logic from the scratch. The Figure 3 and Figure 4 includes the components which consist of the text fields and buttons. Upon submitting the JSP page processed by the JBoss Server to generate the results from the server to display the web page that the candidate profile has been added using the Framework Studio. All the components which were dragged to the Canvas available on the palette and have been added to the web page without writing the code again for the components such as button, text field and labels. All the above said components are reused effectively with developing the components again from the scratch.

## 6. CONCLUSION

The Framework Studio is proposed for the generation of code for a particular domain specific component. The proposed framework is tested for a HR portal application domain. The related components are extracted from the Framework Studio which is integrated with Net Beans IDE for deployment of HR Portal application. This increases the reusability of enterprise applications in particular domain. As more and more components are added to the Framework Studio, many applications can be deployed, which helps in reduction of time and budget. This helps in reduction of risk of software development to a large extent. As per the future work, it is necessary to identify the rarely used components and add to the pallet. One of the issues still lags in the related work is that the metrics for identifying the quality of reusable components. These metrics can be identified in the design phase or in the coding phase can help to reduce the rework by improving quality of reuse of the component and hence improve the productivity due to probabilistic increase in reuse level. So, the developed systems can be used to enhance the productivity and quality of software development much faster.

### ACKNOWLEDGEMENTS

The work was partly supported by the R & D Cell of Muffakham Jah College of Engineering & Technology, Hyderabad, India. The authors would like to thank to all the people from Industry and Academia for their active support.

## Authors

**N Md Jubair Basha** received his B.Tech. (IT) and M.Tech (IT) from JNTUH, Hyderabad He is presently working as Assistant Professor in Department of Information Technology, Muffakham Jah College of Engineering and Technology, Hyderabad, India. His research interest includes Software Reusability, Data Mining and Mobile Computing. He is an active member of IEEE and CSI. You can reach him at nawabjubair@gmail.com.

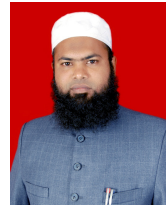

**Dr. Salman Abdul Moiz** is working as a Professor in CSE/IT department at MVSR Engineering College, Hyderabad. He received B.Sc (Electronics) from Osmania University, MCA from Osmania University, M.Tech (CSE) from Osmania University, and M.Phil (CS) from Madurai Kamaraj University and Ph.D (CSE) from Osmania University. He worked as Research Scientist at Centre for Development of Advanced Computing, Bangalore. He has published 31 papers in various National/International Conferences and Journals. His areas of interests include Mobile databases, Software Process Improvements; Component based software development & Disaster Recovery. He is an active member of IEEE, IETE and CSI.

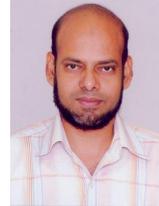